\documentclass[aps,tightenlines,superscriptaddress,twocolumn]{revtex4-1}
\usepackage{graphicx}
\usepackage{bm}
\usepackage{times}
\usepackage{lineno}
\usepackage{amssymb}
\usepackage[section]{placeins}
\usepackage[pscoord]{eso-pic}

\begin{document}

\title{Probing Strangeness Canonical Ensemble with $K^{-}$, $\phi(1020)$ and $\Xi^{-}$ Production in Au+Au Collisions at ${\sqrt{s_{\rm NN}} = \rm{3\,GeV}}$}% Force line breaks with \\

\affiliation{Abilene Christian University, Abilene, Texas   79699}
\affiliation{AGH University of Science and Technology, FPACS, Cracow 30-059, Poland}
\affiliation{Alikhanov Institute for Theoretical and Experimental Physics NRC "Kurchatov Institute", Moscow 117218}
\affiliation{Argonne National Laboratory, Argonne, Illinois 60439}
\affiliation{American University of Cairo, New Cairo 11835, New Cairo, Egypt}
\affiliation{Brookhaven National Laboratory, Upton, New York 11973}
\affiliation{University of California, Berkeley, California 94720}
\affiliation{University of California, Davis, California 95616}
\affiliation{University of California, Los Angeles, California 90095}
\affiliation{University of California, Riverside, California 92521}
\affiliation{Central China Normal University, Wuhan, Hubei 430079 }
\affiliation{University of Illinois at Chicago, Chicago, Illinois 60607}
\affiliation{Creighton University, Omaha, Nebraska 68178}
\affiliation{Czech Technical University in Prague, FNSPE, Prague 115 19, Czech Republic}
\affiliation{Technische Universit\"at Darmstadt, Darmstadt 64289, Germany}
\affiliation{ELTE E\"otv\"os Lor\'and University, Budapest, Hungary H-1117}
\affiliation{Frankfurt Institute for Advanced Studies FIAS, Frankfurt 60438, Germany}
\affiliation{Fudan University, Shanghai, 200433 }
\affiliation{University of Heidelberg, Heidelberg 69120, Germany }
\affiliation{University of Houston, Houston, Texas 77204}
\affiliation{Huzhou University, Huzhou, Zhejiang  313000}
\affiliation{Indian Institute of Science Education and Research (IISER), Berhampur 760010 , India}
\affiliation{Indian Institute of Science Education and Research (IISER) Tirupati, Tirupati 517507, India}
\affiliation{Indian Institute Technology, Patna, Bihar 801106, India}
\affiliation{Indiana University, Bloomington, Indiana 47408}
\affiliation{Institute of Modern Physics, Chinese Academy of Sciences, Lanzhou, Gansu 730000 }
\affiliation{University of Jammu, Jammu 180001, India}
\affiliation{Joint Institute for Nuclear Research, Dubna 141 980}
\affiliation{Kent State University, Kent, Ohio 44242}
\affiliation{University of Kentucky, Lexington, Kentucky 40506-0055}
\affiliation{Lawrence Berkeley National Laboratory, Berkeley, California 94720}
\affiliation{Lehigh University, Bethlehem, Pennsylvania 18015}
\affiliation{Max-Planck-Institut f\"ur Physik, Munich 80805, Germany}
\affiliation{Michigan State University, East Lansing, Michigan 48824}
\affiliation{National Research Nuclear University MEPhI, Moscow 115409}
\affiliation{National Institute of Science Education and Research, HBNI, Jatni 752050, India}
\affiliation{National Cheng Kung University, Tainan 70101 }
\affiliation{Nuclear Physics Institute of the CAS, Rez 250 68, Czech Republic}
\affiliation{Ohio State University, Columbus, Ohio 43210}
\affiliation{Institute of Nuclear Physics PAN, Cracow 31-342, Poland}
\affiliation{Panjab University, Chandigarh 160014, India}
\affiliation{Pennsylvania State University, University Park, Pennsylvania 16802}
\affiliation{NRC "Kurchatov Institute", Institute of High Energy Physics, Protvino 142281}
\affiliation{Purdue University, West Lafayette, Indiana 47907}
\affiliation{Rice University, Houston, Texas 77251}
\affiliation{Rutgers University, Piscataway, New Jersey 08854}
\affiliation{Universidade de S\~ao Paulo, S\~ao Paulo, Brazil 05314-970}
\affiliation{University of Science and Technology of China, Hefei, Anhui 230026}
\affiliation{Shandong University, Qingdao, Shandong 266237}
\affiliation{Shanghai Institute of Applied Physics, Chinese Academy of Sciences, Shanghai 201800}
\affiliation{Southern Connecticut State University, New Haven, Connecticut 06515}
\affiliation{State University of New York, Stony Brook, New York 11794}
\affiliation{Instituto de Alta Investigaci\'on, Universidad de Tarapac\'a, Arica 1000000, Chile}
\affiliation{Temple University, Philadelphia, Pennsylvania 19122}
\affiliation{Texas A\&M University, College Station, Texas 77843}
\affiliation{University of Texas, Austin, Texas 78712}
\affiliation{Tsinghua University, Beijing 100084}
\affiliation{University of Tsukuba, Tsukuba, Ibaraki 305-8571, Japan}
\affiliation{Valparaiso University, Valparaiso, Indiana 46383}
\affiliation{Variable Energy Cyclotron Centre, Kolkata 700064, India}
\affiliation{Warsaw University of Technology, Warsaw 00-661, Poland}
\affiliation{Wayne State University, Detroit, Michigan 48201}
\affiliation{Yale University, New Haven, Connecticut 06520}

\author{M.~S.~Abdallah}\affiliation{American University of Cairo, New Cairo 11835, New Cairo, Egypt}
\author{B.~E.~Aboona}\affiliation{Texas A\&M University, College Station, Texas 77843}
\author{J.~Adam}\affiliation{Brookhaven National Laboratory, Upton, New York 11973}
\author{L.~Adamczyk}\affiliation{AGH University of Science and Technology, FPACS, Cracow 30-059, Poland}
\author{J.~R.~Adams}\affiliation{Ohio State University, Columbus, Ohio 43210}
\author{J.~K.~Adkins}\affiliation{University of Kentucky, Lexington, Kentucky 40506-0055}
\author{G.~Agakishiev}\affiliation{Joint Institute for Nuclear Research, Dubna 141 980}
\author{I.~Aggarwal}\affiliation{Panjab University, Chandigarh 160014, India}
\author{M.~M.~Aggarwal}\affiliation{Panjab University, Chandigarh 160014, India}
\author{Z.~Ahammed}\affiliation{Variable Energy Cyclotron Centre, Kolkata 700064, India}
% ...

\collaboration{STAR Collaboration}\noaffiliation

%\input{star-author-list-2021-06-22.iop.tex}

%\linenumbers

\date{\today}

\begin{abstract}
We report the first multi-differential measurements of strange hadrons of $K^{-}$, $\phi$ and $\Xi^{-}$ yields as well as the ratios of $\phi/K^-$ and $\phi/\Xi^-$ in Au+Au collisions at ${\sqrt{s_{\rm NN}} = \rm{3\,GeV}}$ with the STAR experiment fixed target configuration at RHIC. The $\phi$ mesons and $\Xi^{-}$ hyperons are measured through hadronic decay channels, $\phi\rightarrow K^+K^-$ and $\Xi^-\rightarrow \Lambda\pi^-$. 
Collision centrality and rapidity dependence of the transverse momentum spectra for these strange hadrons are presented. The $4\pi$ yields and ratios are compared to thermal model and hadronic transport model predictions.
At this collision energy, thermal model with grand canonical ensemble (GCE) under-predicts the $\phi/K^-$ and $\phi/\Xi^-$ ratios while the result of canonical ensemble (CE) calculations reproduce $\phi/K^-$, with the correlation length $r_c \sim 2.7$\,fm, and $\phi/\Xi^-$, $r_c \sim 4.2$\,fm, for the 0-10\% central collisions. Hadronic transport models including high mass resonance decays could also describe the ratios.
While thermal calculations with GCE work well for strangeness production in high energy collisions, the change to CE at $\rm{3\,GeV}$ implies a rather different medium property at high baryon density. 
\end{abstract}

\maketitle

%\linenumbers

% Chapter one
\section{Introduction}
\label{introduction}

Relativistic heavy ion physics is aiming at the detailed investigation of phase structures of strongly interacting matter, governed by quantum chromodynamics (QCD), under extreme conditions of high temperature and density~\cite{StarWhitePaper_2005,akiba2015hot,Busza_ARNPS:2018}. Particle production has been studied to investigate properties of the produced QCD matter in heavy-ion collisions. The strange quark mass is comparable to the QCD scale ($\Lambda_{\rm{QCD}}$\,$\sim\textup{200 MeV}$), therefore strange quark plays an important role in studying the QCD phase diagram and the Equation-of-State (EoS), particularly in the high density region~\cite{Rafelski:1982pu,Koch:1986ud,KO.PhysRevLett.55.2661:1985,FUCHS20061_kaons:2006,KO_sQM2017,Ks0_Lambda_HADES:2019}. 

Statistical thermal models have often been used to characterize thermal properties of the produced media~\cite{Rafelski_1980279,Cleymans:1992zc,Braun-Munzinger:1994ewq,Becattini:1997ii,Braun-Munzinger:1999hun,Florkowski:2001fp,Braun-Munzinger:2001hwo,BraunMunzinger:2003zd,Redlich_CE:2004,Rafelski_PRC:2010,Andronic_2018Naure}. In these models, grand canonical ensemble (GCE) and canonical ensemble (CE) statistical descriptions can be applied to conserve electric charge, baryon number, and strangeness number in order to compute the final state particle yields. Both GCE and CE models are able to describe various particle yields including strange particles produced in heavy-ion collisions at RHIC and the LHC at center-of-mass energy ($\sqrt{s_{\rm NN}}$) greater than 7.7\,GeV. It has been argued that at lower energies, strangeness number needs to be conserved locally on an event-by-event basis described by the CE, which leads to a reduction in the yields of hadrons with non-zero strangeness number (``Canonical Suppression")~\cite{Rafelski_1980279,Redlich:2001kb,Rafelski_2002}, but not for the $\phi(1020)$ meson with zero net strangeness number (S=0). The $\phi/K^-$ ratio is expected to increase with decreasing collision energy in models using the CE treatment for strangeness, opposite to the trend in the GCE treatment. The canonical suppression power for $\Xi^-$ (S=2) is even larger than for $K^-$ (S=1). The $\phi/K^-$ and $\phi/\Xi^-$ ratios offer a unique test to scrutinize thermodynamic properties of strange quarks in the hot and dense QCD environment.

In heavy-ion collisions, the near/sub-threshold production of multi-strange hadrons can be achieved from the multiple collisions of nucleons, produced particles, and short-lived resonances~\cite{ZEEB2004297}. The particle production in heavy-ion collisions below its free nucleon-nucleon (NN) threshold ($\sqrt{s_{\rm NN}}$ $\sim$2.89\,GeV for $\phi$ and $\sim$3.25\,GeV for $\Xi^-$) is expected to be sensitive to the stiffness of the nuclear EoS at high density~\cite{yong2021double}, as it is for single-strange hadrons~\cite{KO.PhysRevLett.55.2661:1985,FUCHS20061_kaons:2006}. The near/sub-threshold production further provides the possibility
to observe exotic states of QCD matter~\cite{McLerran:2007qj} and signatures of ``soft deconfinement"~\cite{Fukushima:2020cmk}.
 
Previous measurements 
show that the $\phi/K^-$ ratio in heavy-ion collisions stays remarkably flat ($\sim$0.15) at collision energies ${\sqrt{s_{\rm NN}} > \textup{5 GeV}}$~\cite{E917_phi:2004,NA49_phi:2008,star_bes_strangeness:2020}. 
%Recent measurements of the $\phi/K^-$ ratio in heavy-ion collisions at collision energies below the $\phi$ free NN-threshold %from HADES and FOPI 
%show a hint of relative enhancement compared to those from high energies
%at RHIC and the LHC
At collision energies close to or below the $\phi$ and $\Xi$ NN-thresholds, recent measurements of $\phi/K$ and $\phi/\Xi$ ratios from HADES and FOPI have achieved a significance about 2.2-3.8 sigma in central heavy ion collisions, and the results indicate a relative enhancement compared to those at high energies
~\cite{HADES_phi_ArKCl:2009,FOPI_phi_NiNi:2015,FOPI_phi_AlAl:2016,HADES_phi_AuAu:2018}, indicative of the applicability of the CE description for strangeness production at these energies. 
%Note, the $\phi$ and $K^-$ are also measured in various small systems which shown the absorption and in-medium modification may play a role~\cite{HADES_PRL_pA:2007,HADES_PRL_W_C:2019,ALICE_PRL_pp:2021}. 
Measurements from $\pi$ or proton induced nuclear reactions~\cite{HADES_PRL_pA:2007,HADES_PRL_W_C:2019}
%~\cite{HADES_PRL_pA:2007,HADES_PRL_W_C:2019,ALICE_PRL_pp:2021}
suggest that absorption in cold nuclear matter may play a role in the $K^-$ and $\phi$ production yields in nuclear collision at low energies. 
In this Letter, we report high precision measurement of $\phi/K^-$ and $\phi/\Xi^-$ ratios in Au+Au collisions at ${\sqrt{s_{\rm NN}} = \rm{3\,GeV}}$ from the STAR experiment.

\section{EXPERIMENT AND DATA ANALYSIS }
\label{dataanalysis}

The dataset used in this analysis
was collected under the fixed target (FXT) setup~\cite{Meehan_2016} in the 2018 RHIC run. 
A single beam was provided by RHIC with total energy equal to 3.85 GeV/nucleon and incident on a gold target of thickness 0.25 mm, corresponding to a 1\% interaction probability.
The target is installed inside the vacuum pipe, 2\,cm below the center of the beam axis, and located 200\,cm to the west of the center of the STAR detector. The main detectors used are the Time Projection Chamber (TPC)~\cite{TPC:2003,Meehan_2016}, the Time of Flight (TOF) detector~\cite{TOF:2012,Meehan_2016}, and the Beam-Beam Counter (BBC)~\cite{BBC_Whitten:2008}. The trigger is provided by the signal in the east BBC detector and at least five hits in the TOF detector.
To best utilize the detector band-width, the beam-on-target collision rate was tuned to around 1.5\,kHz, and the pileup contribution to the triggered event is $<1\%$~\cite{STAR:2021fge}. 
Tracking and particle identification (PID) are done using the TPC and TOF. Both the TPC and TOF detectors have full azimuthal coverage within a pseudorapidity range of 0$<$\,$\eta$\,$<$\,1.88 for the TPC and 0$<$\,$\eta$\,$<$\,1.5 for the TOF in FXT mode~\cite{TPC:2003,TOF:2012,Meehan_2016}.
Events are selected with the offline reconstructed collision vertex within 1.5\,cm of the target center along the beam direction. Approximately 2.6$\times 10^{8}$ minimum bias (MB) triggered events passed the selection criteria and are used in this analysis. 

The centrality class is selected using measured charged particle total multiplicity within the TPC acceptance. 
A Monte Carlo Glauber model, used in conjunction with a negative binomial distribution to model particle production in hadronic collisions, is optimized in order to best match the data and determine the centrality class~\cite{MC_Ray_2008,STAR:2021fge}.
%The trigger selection is fully efficient for 0--50\% centrality events and becomes 
Due to the trigger inefficiency in the low multiplicity region (corresponding to the most peripheral collisions), we only report the results from the 0--60\% centrality class in this paper. In addition, in order to reduce the pile-up contamination, events above the reference multiplicity of 195 are removed from the most central centrality class.

$\phi$ mesons are reconstructed via the decay channel $\phi\rightarrow K^+K^-$ with a branching ratio (BR) of ($49.2\pm0.5$)\% , while the $\Xi^{-}$ hyperons decay via $\Xi^-\rightarrow\Lambda\pi^-\rightarrow p\pi^-\pi^-$ with a BR of ($63.8\pm0.5$)\%~\cite{pdg:2020}. $\Xi^-$ reconstruction is performed using the KFParticle package based on the Kalman Filter method~\cite{Kisel:2020lpa,STAR_PRL_Xi_Oemga_polarization:2021}. The charged tracks are reconstructed with the TPC in a 0.5 T uniform magnetic field, and are required to consist of at least 20 TPC hits (out of a maximum of 45) and have a ratio between the number of hit points and the maximum possible number of hit points larger than 0.52 to ensure good tracking and avoid track splitting. %(15 TPC hits required for $\Xi^{-}$ to increase the sample of signal candidates). 
The TPC tracking performance with these requirements in the FXT data is similar to that in other data taken in the collider mode. Monte Carlo simulations also reproduces the distributions of various tracking variables.
The charged tracks are identified via a combination of the ionization energy loss ($dE/dx$) measurement with the TPC and the time-of-flight ($tof$) measurement with the TOF~\cite{Shao:2005iu,Xu:2008th}. 
The resolution-normalized $dE/dx$ or $\beta$ deviation from the expected values are used for the PID selection. 
A minimum $p_T$ cut of 0.2 GeV/$c$ is required in the analysis.
Since the $K^{-}/\pi^{-}$ ratio is much smaller than the $K^{+}/\pi^{+}$ ratio at low energies, to reduce the contamination from $\pi^{-}$ and $e^{-}$ tracks, a strict PID criterion for $K^{-}$ is implemented by requiring the TPC $dE/dx$ and TOF $\beta$ to be within three standard deviations of the expected values. $K^{+}$ tracks used for the $\phi$ analysis are selected with a hybrid algorithm, in which the TPC $dE/dx$ requirement is applied at low momentum $p<0.5$\,GeV/$c$ while an additional TOF $\beta$ requirement is imposed at $p>0.5$\,GeV/$c$. In the $\Xi$ analysis, proton and $\pi^-$ tracks are identified by requiring the TPC $dE/dx$ to be within three standard deviations of the expect values and the TOF $\beta$ requirement is only applied when there is a valid measurement.

\begin{figure}
\centering
\hspace*{-4mm}
\includegraphics[width=0.50\textwidth]{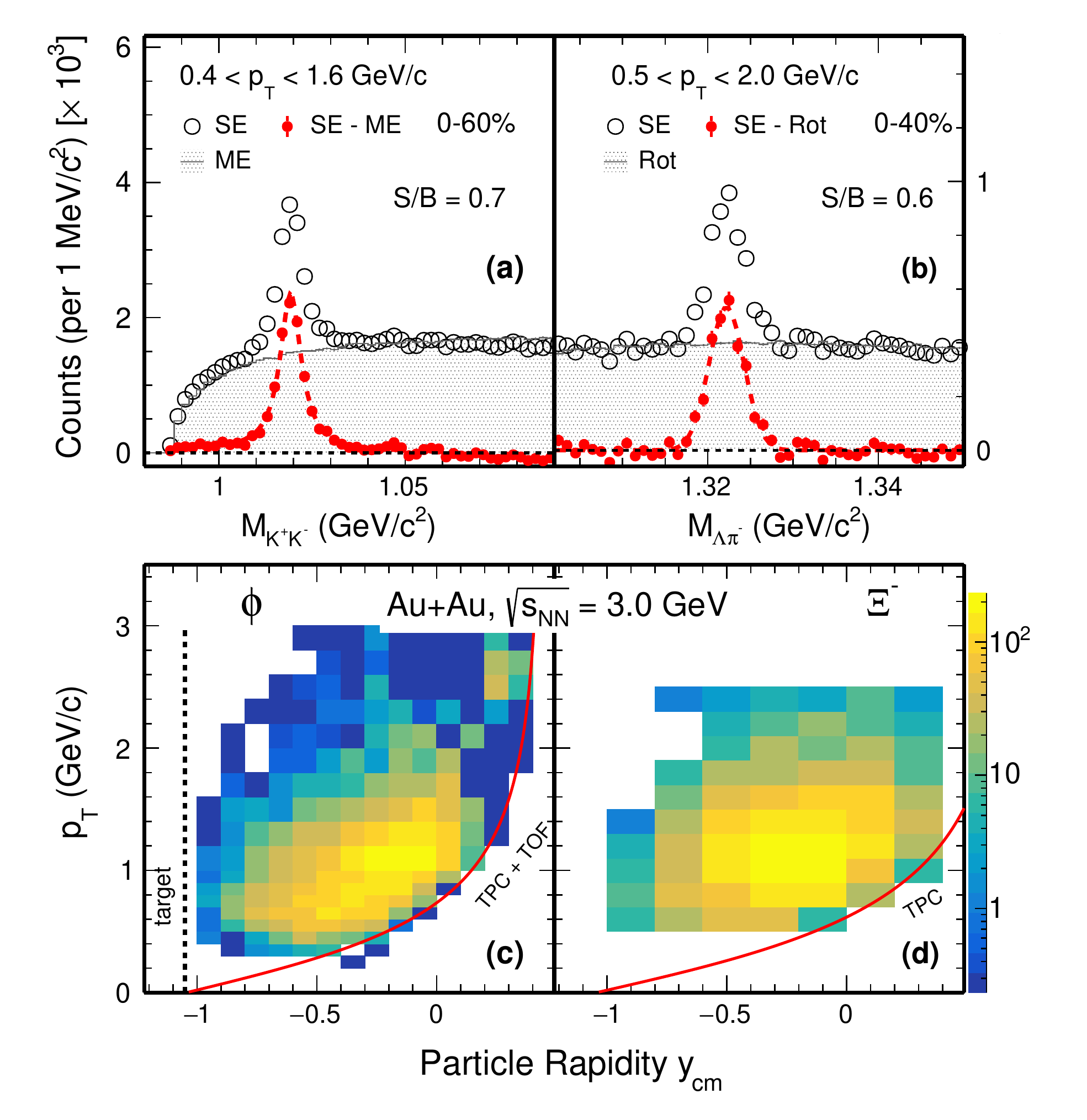}
  \caption{Invariant mass distributions of $K^+K^-$ (a) and  $\Lambda\pi^-$ (b) in Au+Au collisions at ${\sqrt{s_{\rm NN}} = \rm{3\,GeV}}$. Black open circles represent the same-event unlike-sign distribution. The grey shaded histogram represents the normalized mixed-event (rotating daughters for $\Xi^-$) unlike-sign distribution that is used to estimate the combinatorial background. The red solid circles depict the $\phi$ meson (a) and $\Xi^-$ (b) signals obtained by subtracting the combinatorial background from the same-event distribution. Reconstructed $\phi$ (c) and $\Xi^-$ (d) acceptance, $p_T$ vs. rapidity in the center-of-mass frame ($y_{\rm cm}$) in the same collisions. The dotted line indicates the target rapidity location. The red curve represents the TPC and TOF acceptance edge.}
\label{fig:phiSignal} 
\end{figure}

Figure~\ref{fig:phiSignal} (a) shows the invariant mass distribution of $K^+K^-$ pairs in the transverse momentum ($p_{T}$) region of 0.4--1.6 GeV/$c$ for 0--60\% central collisions. The combinatorial background is estimated with the mixed-event (ME) technique in which $K^+$ and $K^-$ from different events of similar characteristics (centrality, event plane angle) are paired. The mixed-event spectra are normalized to the same-event (SE) distributions in the mass range of 1.04--1.08\,GeV/$c^2$. After the subtraction of the combinatorial background, the remainder distribution, shown as red solid circles, 
is fitted with a Breit-Wigner function for the signal plus a linear function which represents the remaining correlated background ($< 1\%$) from a partial reconstruction of strange hadrons. The $\phi$ meson raw yields are extracted from the Breit-Wigner function fit within the corresponding $\pm$3$\Gamma$ mass window (mean value $\mu\sim$ 1.0194\,GeV/$c^2$, full-width-at-half-maximum $\Gamma\sim$ 6.5\,MeV/$c^2$). 
The extracted $\phi$ signal shape is consistent with its intrinsic properties convoluted with the detector smearing effect due to finite momentum resolution ($<3\%$ for single track).
Note that a Voigt function has been used to extract the signal counts as a cross check, and the extracted yields are consistent with the default value within uncertainties. 
Figure~\ref{fig:phiSignal} (b) shows the invariant mass distribution of $\Lambda(p\pi^-)\pi^-$ in the $p_{T}$ region of 0.5--2.0 GeV/$c$ for $\textup{0--40\%}$ central collisions. The combinatorial background is estimated with the rotating daughter (Rot) method, in which a daughter track of $\Xi^-$ is rotated by a random angle between 150 to 210 degrees in the transverse plane. The rotated spectra are normalized to the same-event distributions in the mass ranges of 1.30--1.31 and 1.34--1.35\,GeV/$c^2$. After the combinatorial background is subtracted, the $\Lambda\pi^-$ invariant mass distribution is fitted with a Gaussian for the signal plus a linear function for the remaining correlated background ($< 1\%$). The $\Xi^-$ raw yields are obtained via histogram bin counting from the invariant mass distributions with all background subtracted within mass windows of $\pm$3$\sigma$ ($\mu\sim$ 1.3222\,GeV/$c^2$, Gaussian width $\sigma\sim$\,2.0\,MeV/$c^2$). The signal-to-background (S/B) ratio for $\phi$ and $\Xi^-$ within the reconstructed mass windows is about 0.7 and 0.6 respectively. The reconstructed $\phi$ and $\Xi^-$ acceptances ($p_T$ vs. $y_{cm}$) in the collision center-of-mass frame are shown in Fig.~\ref{fig:phiSignal} (c) and (d), respectively.
The target is located at $y_{cm} = -1.05$, using the convention where the beam travels in the positive direction. The red curve represents the TPC and TOF acceptance edge. %The reconstructed $\phi$ and $\Xi$ in this analysis cover the range from the target to mid-rapidity.

\begin{figure*}
\centering
\hspace*{-10mm}
\includegraphics[width=0.43\textwidth]{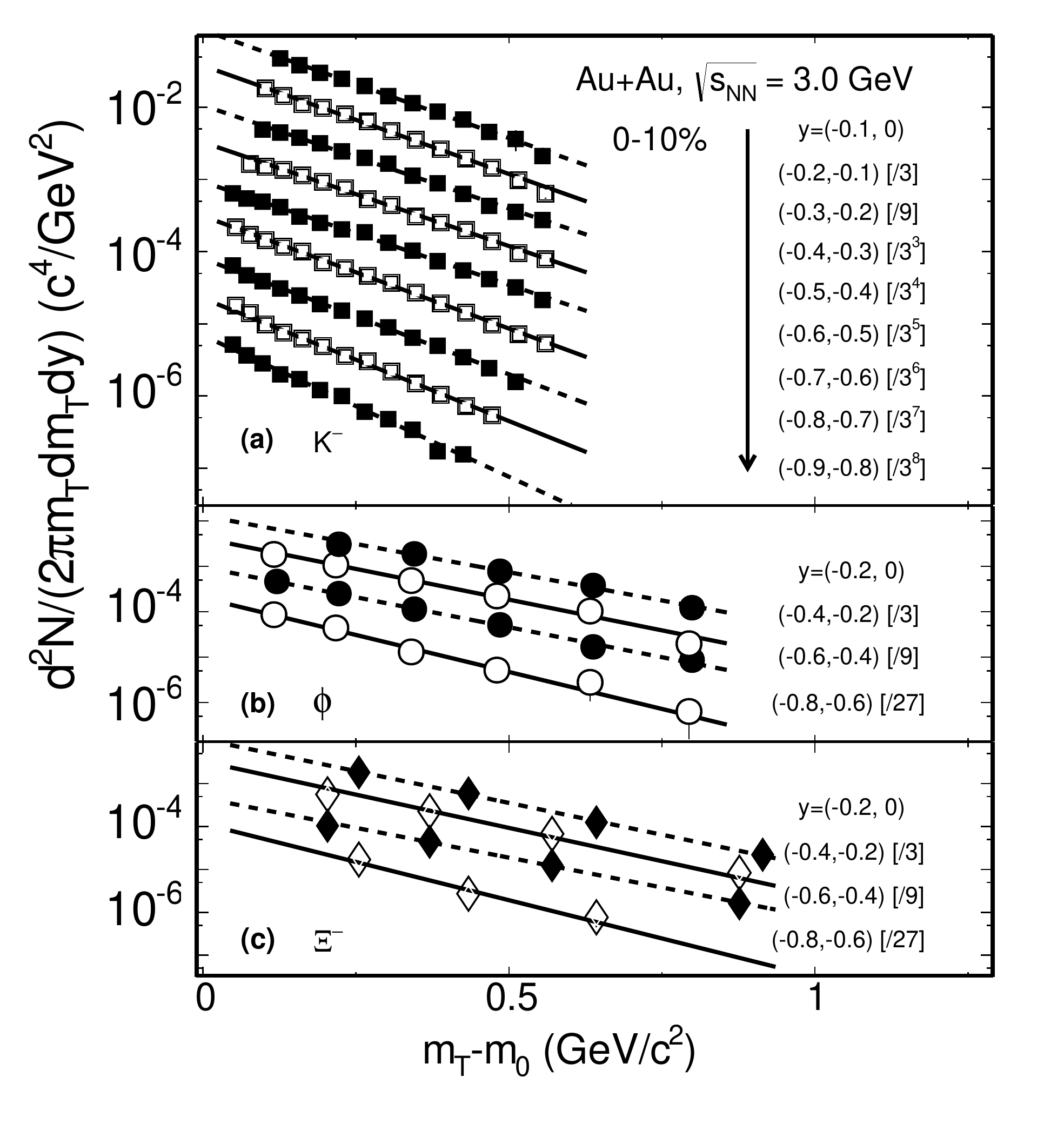}
\hspace*{+10mm}
\includegraphics[width=0.43\textwidth]{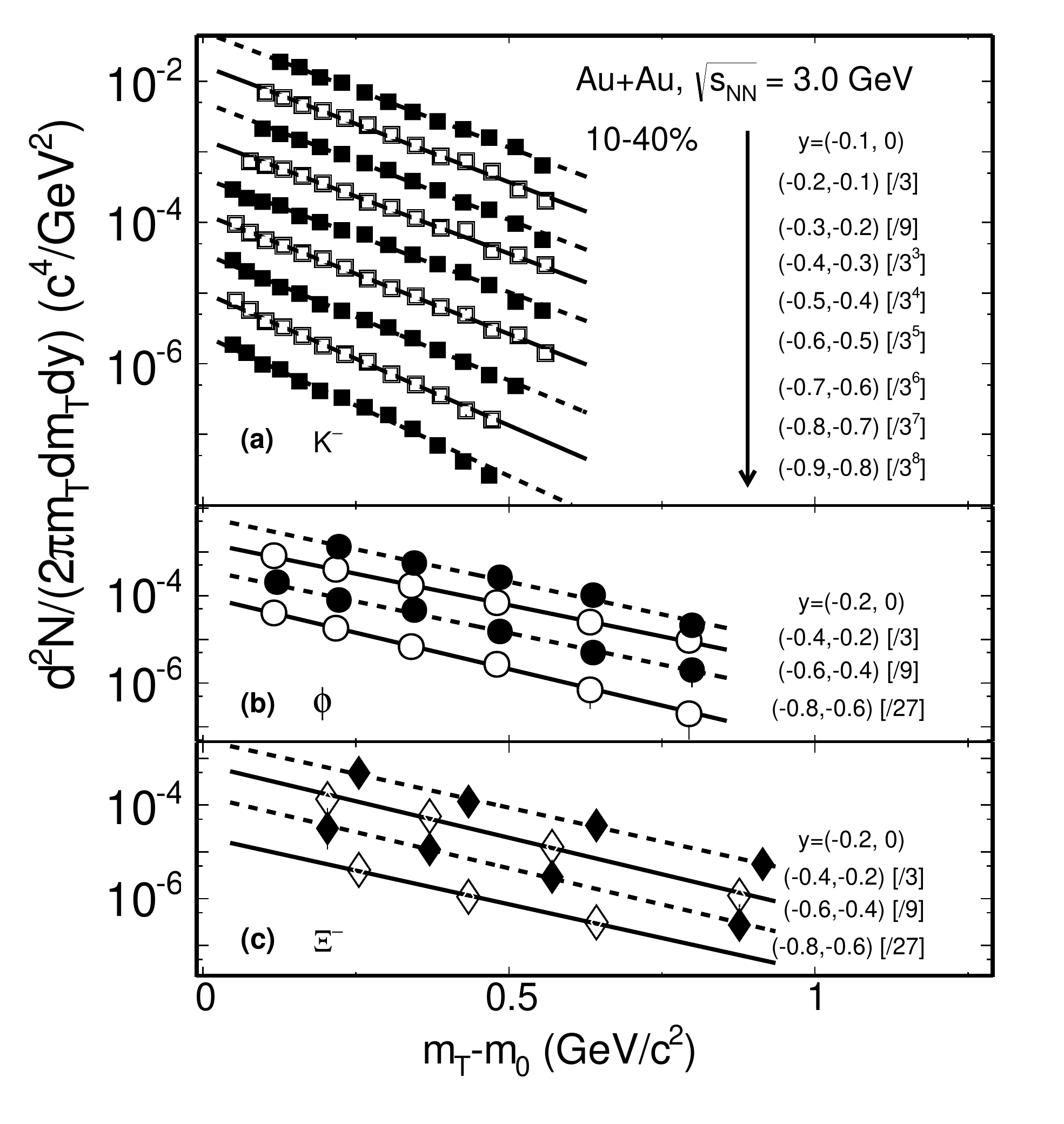}
\caption{$K^-$ (a), $\phi$ meson (b) and $\Xi^-$ (c) invariant yields as a function of $m_T-m_0$ for various rapidity regions in 0--10\% (left) and  10--40\% (right) centrality Au+Au collisions at ${\sqrt{s_{\rm NN}} = \rm{3\,GeV}}$. Statistical and systematic uncertainties are added quadratically here for plotting. Solid and dashed black lines depict $m_T$ exponential function fits to the measured data points with scaling factors in each rapidity windows.}
\label{fig:phimTSpectra21}
\end{figure*}

Particle raw yields are calculated in each centrality and $p_{T}$ bin within each rapidity slice. 
The raw yields are corrected for the TPC acceptance and tracking efficiency, %$\varepsilon_{\rm TPC}$, 
the particle identification efficiency, %$\varepsilon_{\rm PID}$ 
and the TOF matching and PID efficiency. 
The TPC acceptance and tracking efficiency is obtained using the standard STAR embedding technique~\cite{Adamczyk:2017iwn,star_bes_strangeness:2020}, in which a small number of MC tracks are processed through the GEANT (v3.21) simulation~\cite{GEANT3:1994}, then mixed with the real data and reconstructed using the same algorithm as in the real data. The TPC PID, TOF matching and PID efficiencies are obtained from the data-driven method similar as in Ref.~\cite{STAR:HFTD0:2018zdy}.
The final average reconstruction (including acceptance etc.) efficiency is $\sim$0.30, 0.04, and 0.02 for $K^-$, $\phi$ and $\Xi^-$, respectively. MC embedding simulation also reproduces various topological variables used in the $\Xi^-$ reconstruction. As a cross-check, we conducted the measurement of $\Xi^{-}$ lifetime from the same data and the result is $164.2\pm6.6$ (stat.)\,ps, consistent with the PDG value, $163.9\pm1.5$\,ps. The corrected $p_T$ spectra in symmetric rapidity bins (-0.2,0) vs. (0,0.2) are also consistent.

The systematic uncertainty of the raw yield extraction is estimated by changing the histogram fitting method to bin counting method or by changing the fitting ranges. The maximum difference between these scenarios and the default one is considered as one standard deviation. The contribution varies by $p_T$, rapidity, and centrality and the overall contribution is less than 5\% for the invariant yield. The systematic uncertainty in the TPC acceptance and efficiency correction $\varepsilon_{\rm TPC}$ is estimated by varying the cuts on track selection criteria ~\cite{Adamczyk:2017iwn} and topological variables (for $\Xi^-$ only). The contribution to the total yield is 4-5\% for $K^-$, 13-16\% for $\phi$ and 6-10\% for $\Xi^-$. This leads to a 10-13\% (12-18\%) uncertainty in the measured $\phi/K^-$ ($\phi/\Xi^-$) ratio. The uncertainty of the PID efficiency correction is estimated by varying the PID selection cuts and the contribution is less than 3\% to the total yield.
For the $p_T$ integrated yield, the uncertainty due to the extrapolation to the full $p_T$ range is estimated by choosing several fitting functions including Levy, Blast-Wave, $m_T$-exponential, $p_T$-exponential~\cite{STAR_particleYield:2009}, and the maximum difference between these scenarios and the default one ($m_T$-exponential) is quoted as one standard deviation. This contribution is 5-7\% for $K^-$, 14-17\% for $\phi$ and 13-15\% for $\Xi^-$, respectively. 
This measurement covers nearly the full rapidity range from $y$=0 to the target region. The systematic uncertainty due to the rapidity coverage extrapolation is negligible compared to other systematic sources.
For each individual $\phi$-meson, $K^-$ and $\Xi^-$ measurement, some of the uncertainties are correlated or partially correlated (e.g. TPC and PID). To avoid the correlation in the ratio measurement, we vary the above cuts simultaneously for $\phi$, $K^-$ and $\Xi^-$, then quote the difference in the final ratios as the systematic uncertainties.

\section{RESULTS AND DISCUSSIONS}
\label{results}

\begin{figure}
\centering
\hspace*{-10mm}
\includegraphics[width=0.43\textwidth]{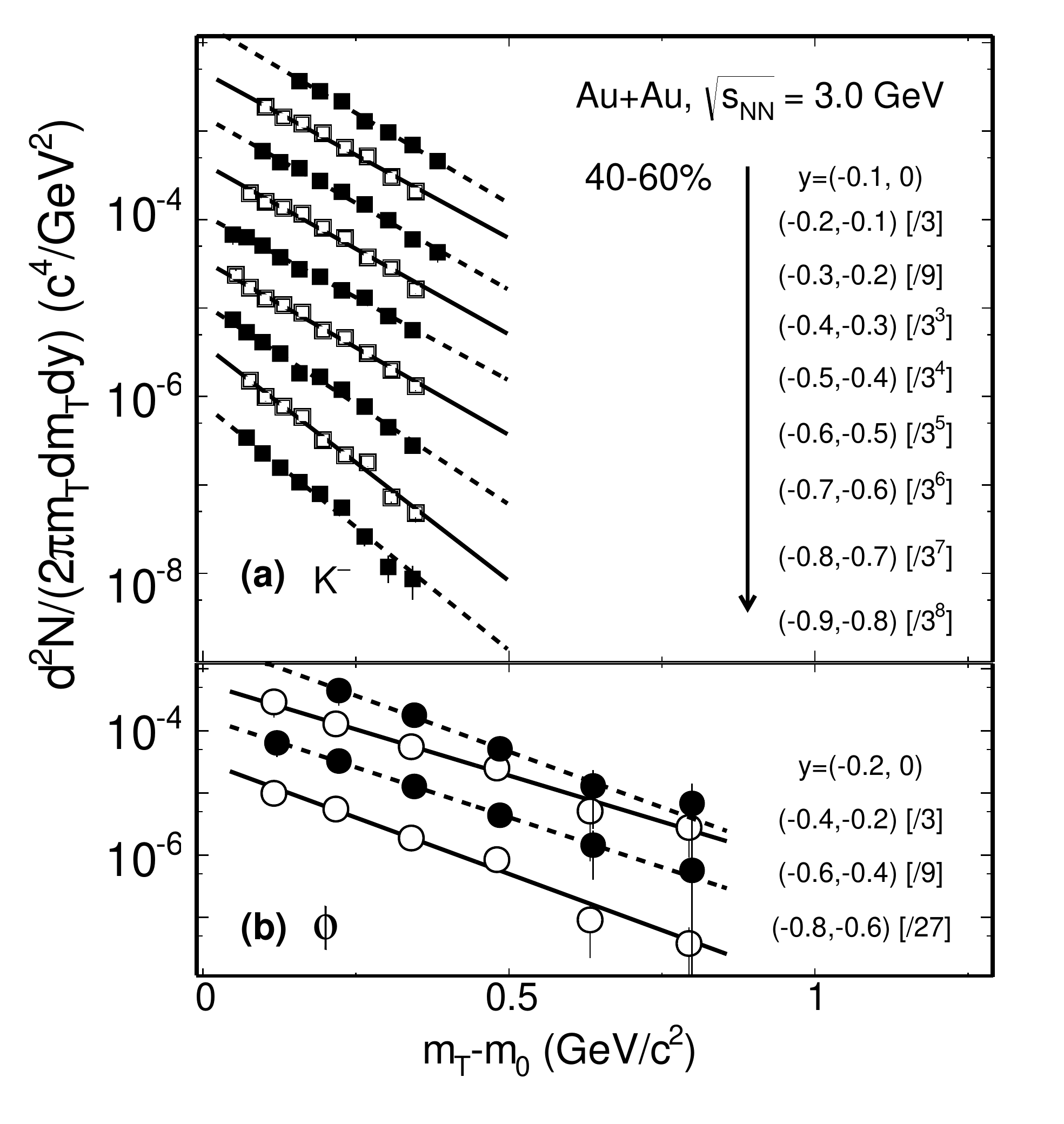}
\caption{$K^-$ (a) and $\phi$ meson (b) invariant yields as a function of $m_T-m_0$ for various rapidity regions in 40--60\% centrality Au+Au collisions at ${\sqrt{s_{\rm NN}} = \rm{3\,GeV}}$.}
\label{fig:phimTSpectra23}
\end{figure}

Figure~\ref{fig:phimTSpectra21} and ~\ref{fig:phimTSpectra23} show the acceptance $\times$ efficiency corrected $K^-$, $\phi$ and $\Xi^-$ invariant yields as a function of $m_T-m_0$ ($m_T = \sqrt{m_{0}^{2}+p_{T}^2/c^{2}}$, where $m_0$ is particle rest mass, and $c$ is the speed of light) for various rapidity ranges in 0--10\%, 10--40\% and 40--60\% Centrality Au+Au collisions at ${\sqrt{s_{\rm NN}} = \rm{3\,GeV}}$. 
Dashed and solid lines depict fits to the spectra with the $m_T$-exponential function in order to extrapolate to the unmeasured $p_T$ ranges ($\sim$20-40\% for $K^-$ which vary rapidity, $\sim$33-50\% for $\phi$ and $\sim$40-60\% for $\Xi^-$). 
The fitted inverse slope parameters indicate harder spectra for the $\phi$-mesons compared to the $K^-$ and $\Xi^-$ within uncertainties. The inverse slope parameters gradually decrease from mid-rapidity to forward/backward rapidity and follow the $T_{\rm eff}/\cosh(y)$ distribution well. The inverse slope parameter at $y=0$, $T_{\rm eff}$, is extracted to be $177\pm5 (stat)\pm8 (sys)$\,MeV for $\phi$ meson, $158\pm3 (stat)\pm3 (sys)$\,MeV for $K^{-}$ and $156\pm3 (stat)\pm24 (sys)$\,MeV for $\Xi^{-}$ in 0--10\% central collisions. This agrees with the collision energy dependence trend from other experiments~\cite{NA49_phi:2008,HADES_phi_AuAu:2018}. Table~\ref{table:Teff} lists the extracted $T_{\rm eff}$ parameter for these particles in different centrality bins from this measurement.

\begin{table*}
\centering{
  \caption{Inverse slope parameter $T_{\rm eff}$ at $y=0$ for the $m_T$ spectra of $\phi$, $K^-$, $\Xi^-$ in Au+Au collisions at ${\sqrt{s_{\rm NN}} = \rm{3\,GeV}}$. The first error given corresponds to the statistical one, the second to the systematic error.}
\begin{tabular}{c|c|c|c} \hline \hline
  Centrality & $\phi$ $T_{\rm eff}$ (MeV) & $K^-$ $T_{\rm eff}$ (MeV) & $\Xi^-$ $T_{\rm eff}$ (MeV)    \\ \hline
  ~~~0--10\%~~~   & ~~~$177\pm5\pm8$~~~  & ~~~$158\pm3\pm3$~~~  & ~~~$156\pm3\pm24$~~~ \\
  10--40\%  & $159\pm4\pm5$  & $142\pm3\pm3$ & $146\pm4\pm17$ \\
  40--60\%  & $151\pm5\pm11$  & $115\pm4\pm 4$  & ---  \\  \hline \hline
\end{tabular}
\label{table:Teff}
}
\end{table*}

\begin{figure}
\centering
\hspace*{-4mm}
\includegraphics[width=0.52\textwidth]{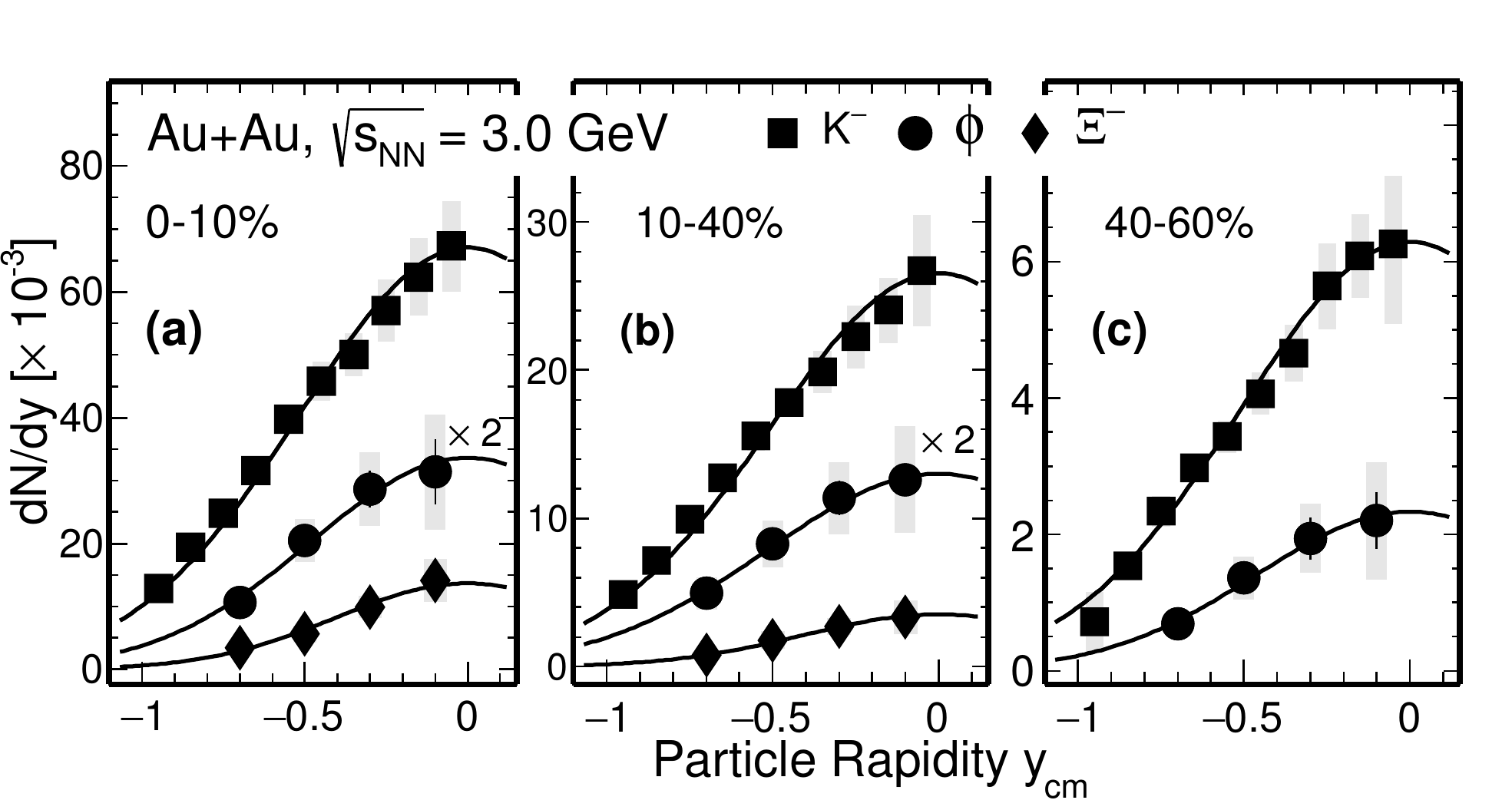}
  \caption{Rapidity density distributions of $K^-$ (squares), $\phi$ meson (circles) and $\Xi^-$ (diamonds) $p_T$-integrated yields $dN/dy$ in 0--10\% (a), 10--40\% (b) and 40--60\% central (c) Au+Au collisions at ${\sqrt{s_{\rm NN}} = \rm{3\,GeV}}$.
  Solid lines depict Gaussian function fits to the data points.}
\label{fig:phiYSpectra} 
\end{figure}

The $p_T$ integrated rapidity distributions $dN/dy$ are displayed in Fig.~\ref{fig:phiYSpectra} for Au+Au collisions at ${\sqrt{s_{\rm NN}} = \rm{3\,GeV}}$ for three different centralities. 
Solid curves depict Gaussian function fits to the data points with the centroid parameter fixed to zero. They are used to extrapolate to the unmeasured rapidity region ($\sim$5\% for $K^-$, $\sim$9\% for $\phi$ and $\sim$6\% for $\Xi^-$) for calculating total multiplicities. 
The integral yields follow the collision energy trend from other experiments and drop quickly toward the low energies around threshold  ~\cite{NA49_phi:2008,HADES_phi_ArKCl:2009,FOPI_phi_NiNi:2015,FOPI_phi_AlAl:2016,HADES_phi_AuAu:2018}.

% \begin{figure}
% \centering
% \hspace*{-4mm}
% \includegraphics[width=0.43\textwidth]{fig2_mT_spectra-eps-converted-to.pdf}
%   \caption{$K^-$ (a), $\phi$ meson (b) and $\Xi^-$ (c) invariant yields as a function of $m_T-m_0$ for various rapidity regions in 0--10\% central Au+Au collisions at ${\sqrt{s_{\rm NN}} = \rm{3\,GeV}}$. Statistical and systematic uncertainties are added quadratically here for plotting. Solid and dashed black lines depict $m_T$ exponential function fits to the measured data points with scaling factors in each rapidity windows.}
% \label{fig:phimTSpectra} 
% \end{figure}

\begin{table*}
\centering{
  \caption{$\phi$, $K^-$, $\Xi^-$ integrated yields, $T_{\rm eff}$ and $\phi/K^-$ and $\phi/\Xi^-$ ratios for given centrality classes in Au+Au collisions at ${\sqrt{s_{\rm NN}} = \rm{3\,GeV}}$. The first error given corresponds to the statistical one, the second to the systematic error.}
\begin{tabular}{c|c|c|c|c|c} \hline \hline
  Centrality & $\phi$ $(10^{-3})$  & $K^-$ $(10^{-2})$ & $\phi/K^-$ & $\Xi^-$ $(10^{-3})$ & $\phi/\Xi^-$  \\ \hline
  ~~~0--10\%~~~   & ~~$20.1\pm1.4\pm3.8$~~  & ~~$8.70\pm0.02\pm 0.53$~~  & ~~$0.231\pm0.016\pm0.042$~~ & ~~$13.9\pm0.8\pm2.4$~~ & ~~$1.45\pm0.13\pm0.34$~~ \\
  10--40\%  & $8.5\pm0.4\pm1.7$  & $3.39\pm0.01\pm 0.20$  & $0.249\pm0.011\pm0.046$ & $3.61\pm0.32\pm0.59$ & $2.34\pm0.23\pm0.65$ \\
  40--60\%  & $2.6\pm0.2\pm0.5$  & $0.79\pm0.01\pm 0.06$  & $0.327\pm0.029\pm0.069$ & --- & --- \\ \hline \hline
\end{tabular}
\label{table:yieldTratio}
}
\end{table*}

\begin{figure}
\centering
\hspace*{-4mm}
\includegraphics[width=0.38\textwidth]{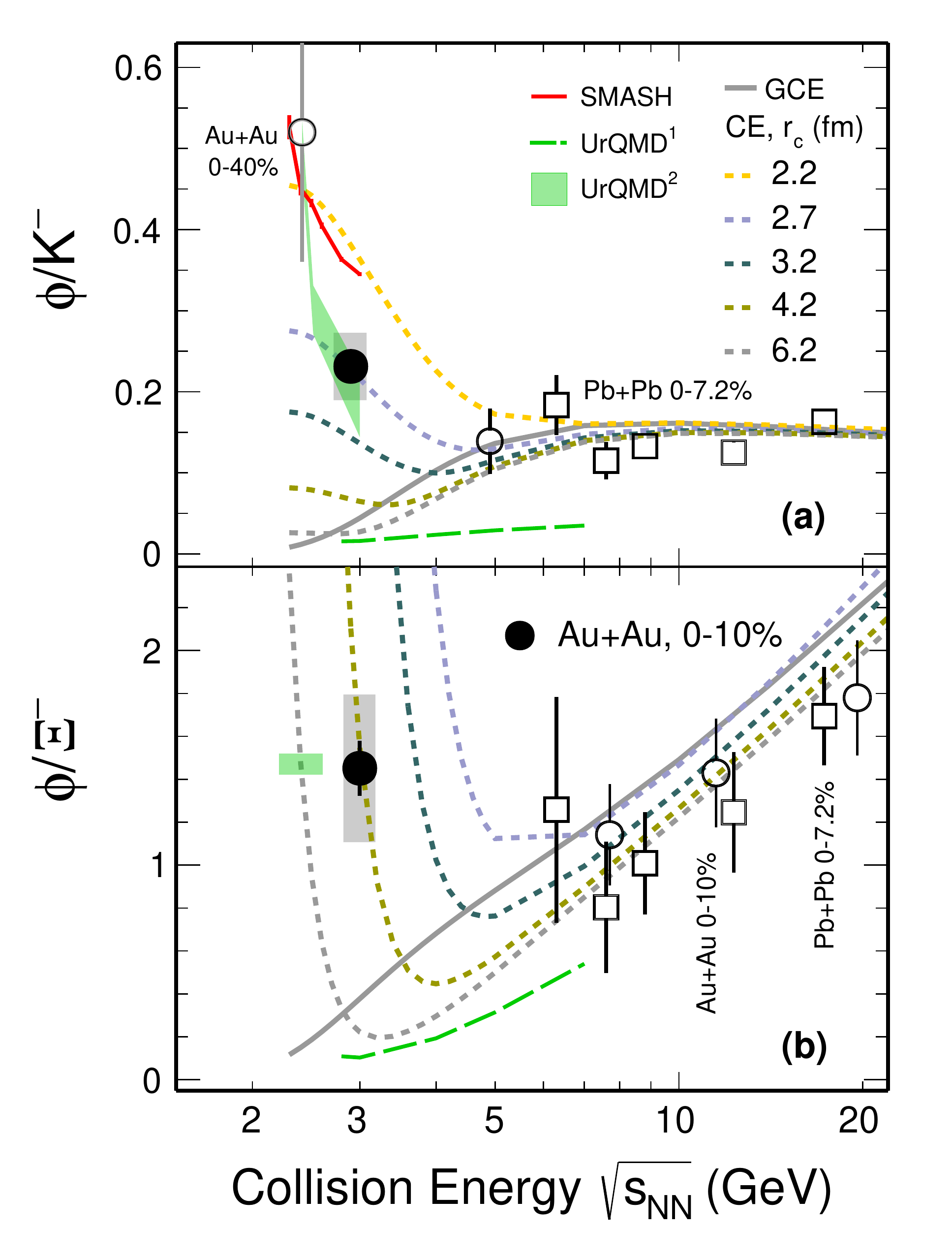}
  \caption{$\phi/K^-$ (a) and $\phi/\Xi^-$ (b) ratio as a function of collision energy, $\sqrt{s_{\rm NN}}$. The solid black circles show the measurements presented here in 0-10\% centrality bin, while empty markers in black are used for data from various other energies and/or collision systems~\cite{E917_phi:2004,NA49_phi:2008,HADES_phi_ArKCl:2009,Xi_ArKCl_HADES:2009,FOPI_phi_NiNi:2015,FOPI_phi_AlAl:2016,HADES_phi_AuAu:2018,star_bes_strangeness:2020}. The vertical grey bands on the data points represent the systematic uncertainties. The grey solid line represents a THERMUS calculation based on the Grand Canonical Ensemble (GCE) while the dotted lines depict calculations based on the Canonical Ensemble (CE) with different values of the strangeness correlation radius ($r_c$)~\cite{THERMUS_WHEATON200984,Andronic_2018Naure}. The green dashed line, green shaded band and the solid red line show transport model calculations from the public versions $\textup{UrQMD}^{1}$~\cite{UrQMD_2:1998,urQMD:1999}, modified $\textup{UrQMD}^{2}$~\cite{Steinheimer_2015_UrQMD} and SMASH~\cite{Elfner_SMASH:2019}, respectively.}
\label{fig:phi2Kratio} 
\end{figure}

The $\phi/K^-$ and $\phi/\Xi^-$ ratios are presented in Fig.~\ref{fig:phi2Kratio} as a function of collision energy $\sqrt{s_{\rm NN}}$, including the midrapidity data in central Au+Au or Pb+Pb data from the AGS, SPS and RHIC BES at higher energies and $4\pi$\,acceptance data from SIS at lower energies. The black solid circles show our measurements in the 0-10\% centrality bin in Au+Au collisions at ${\sqrt{s_{\rm NN}} = \rm{3\,GeV}}$. The measured $\phi$, $K^-$ and $\Xi^-$ yields in 4$\pi$ and the $\phi/K^-$, $\phi/\Xi^-$ ratios in different centrality bins are listed in Tab.~\ref{table:yieldTratio}. The $\phi/K^-$ and $\phi/\Xi^-$ ratios measured at 3\,GeV are slightly higher than, or comparable to, the values at high energies for $\sqrt{s_{\rm NN}} \geqslant$ 5\,GeV~\cite{NA49_piK2:2002,E917_phi:2004,NA49_phi:2008,NA49_piK:2008,NA49_Xi:2008,STAR_phi_64a200GeV:2009,Xi_ArKCl_HADES:2009,ALICE_phi_2p7TeV:2015,star_bes_strangeness:2020} despite the collision energy being very close to the $\phi$ threshold and below the $\Xi^-$ threshold in NN collisions. Note that the enhancement of $\phi/K^-$ and $\phi/\Xi^-$ were also observed at lower collision energies in ${\sqrt{s_{\rm NN}} = \rm{2.4\,GeV}}$ Au+Au~\cite{HADES_phi_AuAu:2018} and  ${\sqrt{s_{\rm NN}} = \rm{2.6\,GeV}}$ Ar+KCl collisions~\cite{HADES_phi_ArKCl:2009,Xi_ArKCl_HADES:2009}, respectively.

Various curves in Fig.~\ref{fig:phi2Kratio} represent the predictions of $\phi/K^-$ and $\phi/\Xi^-$ ratios from several model calculations in central A+A collisions. Statistical model calculations, based on the Grand Canonical Ensemble and Canonical Ensemble for strangeness with several different choices of strangeness correlation length ($r_c$), were calculated using the $\textup{THERMUS}$ package~\cite{THERMUS_WHEATON200984} with energy dependent freeze-out parameters (chemical freeze-out temperature $T_{\rm ch}$, baryon chemical potentials $\mu_B$) taken from~\cite{Andronic_2018Naure}, for instance, $T_{\rm ch}$ = 72.9\,MeV and $\mu_B$ = 701.4\,MeV for $\sqrt{s_{\rm NN}}$ = 3\,GeV.
We noted that the $\phi/K^-$ and $\phi/\Xi^-$ ratios from GCE depend on strangeness chemical potential, $\mu_{S}$.
From the results of the thermal model fit to the STAR BES-I data~\cite{star_bes_strangeness:2020}, there is an empirical relation $\mu_{S}$ = $\mu_{B}/4$ in the collision energy region between 7.7 - 39\,GeV. The same relation was assumed and used in the GCE calculation at lower energies presented in Fig.~\ref{fig:phi2Kratio}. 
%The same relation, $\mu_{S}$ = $\mu_{B}/4$ from the STAR BES-I data~\cite{star_bes_strangeness:2020}, was used in this GCE calculation. 
%With unprecedented precision, our data exclude the GCE calculation, 
Our measured $\phi/K^-$ and $\phi/\Xi^-$ ratios are larger than this GCE calculation: $\chi^2$/ndf = 26.0/2 ($p$-value $<$ $1e^{-5}$), 
which indicates the 
event-by-event 
strangeness conservation 
is crucial~\cite{BraunMunzinger:2003zd} in such collisions.
The exact GCE calculation depends on the precise determination of $T_{\rm ch}$, $\mu_B$, $\mu_S$ etc, which can be extracted through a global fit to various other particle yields at 3\,GeV.
In the canonical approach, the correlation length, $r_c$, defines a region of the particle production phase space inside which the production of the strangeness is canonically conserved. Both the $\phi/K^-$ and $\phi/\Xi^-$ data from our measurement favor the CE thermodynamics for strangeness with a small strangeness correlation length ($r_c \sim 2.7$\,fm for $\phi/K^-$ and $r_c \sim 4.2$\,fm for $\phi/\Xi^-$). It is worthwhile to point out that the CE calculations with the same $r_c$ parameter cannot describe our $\phi/K^-$ and $\phi/\Xi^-$ data simultaneously. The CE calculation with $r_c \sim 4.2$\,fm describes $\phi/\Xi^-$ well while it deviates by about 3.5$\sigma$ for $\phi/K^-$. %This was also observed in lighter systems and at lower collision energies ~\cite{HADES_phi_ArKCl:2009,Xi_ArKCl_HADES:2009}.
$r_c$ is an approximation in the CE for reproducing the strange production in heavy-ion collisions. It is unclear if the same value of $r_c$ should fit for both S=1 (e.g. Kaon) and S=2 (e.g. $\Xi^-$). On the other hand, transport model calculations~\cite{Steinheimer_2015_UrQMD,Elfner_SMASH:2019} with high mass strange resonances reproduce the data implying that the feed down is relevant.

%The CE implementation in most thermal models introduces the correlation length or volume as an approximation to scale down the GCE yields~\cite{BraunMunzinger:2003zd,THERMUS_WHEATON200984}. The simultaneous description for various (multi-)strange hadron yields in low energies is yet to be tested. Our data, in comparison to the available thermal model calculations, may indicate the need of revisiting the exact implementation of the canonical statistics in these calculations. In addition, as suggested from the comparison to transport model calculations discussed in the following paragraph, the feed-down contributions from baryon resonances to the measured $\phi$ and $\Xi$ may be important and carefully checked out in thermal model calculations in these low energies.
%Eventually, a global thermal model fit with all the particle yields at 3\,GeV will help to precisely determine these thermal parameters in the future, and a consistent picture is expected to reasonable describe all the measured strange hadrons yields simultaneously. 
%A global thermal model fit with all the particle yields at 3\,GeV will help to precisely determine these thermal parameters in the future, and a consistent $r_c$ value is expected to reasonable describe all the measured strange hadrons yields simultaneously. 

%\begin{figure}
%\centering
%\hspace*{-4mm}
%\includegraphics[width=0.35\textwidth]{ratio_vs_Npart.pdf}
%  \caption{$\phi/K^-$ ratio as a function of collision centrality, $N_{part}$. The solid black circles show the measurements presented here in 3 centrality bins, while empty markers in grey are used for data from various other similar energies and/or collision systems~\cite{ANKE_phi:2008,HADES_phi_ArKCl:2009,FOPI_phi_NiNi:2015,FOPI_phi_AlAl:2016,HADES_phi_AuAu:2018}. The vertical grey bands on the data points represent the systematic uncertainties.}
%\label{fig:phi2KratioCent} 
%\end{figure}

Previous measurements from smaller collision systems (Ar+KCl and Al+Al collisions) show comparable or higher $\phi/K^-$ and/or $\phi/\Xi^-$ ratios at energies below 3\,GeV~\cite{HADES_phi_ArKCl:2009,Xi_ArKCl_HADES:2009,FOPI_phi_NiNi:2015,FOPI_phi_AlAl:2016}. The exclusive measurement in $p$+$p$ collisions at $\textup{2.7 GeV}$ shows a much larger $\phi/K^-$ ratio ($1.04\pm0.23$)~\cite{ANKE_phi:2008}, while the measured ratio at $\textup{17.3 GeV}$ ($0.11\pm0.01$)~\cite{NA61SHINE_pp_piKp:2017,NA61SHINE_pp_phi:2020} is comparable to that in central Au+Au/Pb+Pb collisions at similar energies~\cite{NA49_phi:2008,star_bes_strangeness:2020}. The $\phi/\Xi^-$ ratio in $p$+$p$ collisions at 17.3\,GeV~\cite{NA61SHINE_pp_phi:2020,NA61SHINE_pp_Xi:2020}, $5.09\pm0.36$, is also significantly larger than that in central Au+Au/Pb+Pb collisions ~\cite{NA49_phi:2008,NA49_Xi:2008,star_bes_strangeness:2020}. In our measurement at 3 GeV, there is no obvious difference in the $\phi/K^-$ ratio between the \textup{0--10\%} and \textup{10--40\%} central bins, while the result in the most peripheral 40--60\% central bin shows a hint of a larger value, as shown in Tab.~\ref{table:yieldTratio}. Similarly, the $\phi/\Xi^-$ ratio in mid-central collisions seems to be larger than that in central collisions. Overall, these observations are qualitatively consistent with the expectation that a smaller canonical volume in the smaller system leads to a higher observed $\phi/K^-$ and/or $\phi/\Xi^-$ ratio.

Hadronic transport models are widely used in the high baryon density region to study the properties of the produced dense matter~\cite{UrQMD_2:1998,urQMD:1999,Hartnack:2011cn,Steinheimer_2015_UrQMD,Elfner_SMASH:2019,Song:2020clw}. In the modified version of the Ultra-relativistic Quantum Molecular Dynamics (UrQMD) model~\cite{Steinheimer_2015_UrQMD}, $\textup{UrQMD}^{2}$, new decay channels from high mass baryon resonances to $\phi$ and $\Xi^-$ are deployed. The relevant decay branching fraction was determined by fitting the experimental data from $p$+$p$ collisions~\cite{ANKE_phi:2008}. From the comparison shown in Fig.~\ref{fig:phi2Kratio}, the modified $\textup{UrQMD}^{2}$ calculation for central ($\rm{b}<5\,\rm{fm}$) Au+Au collisions agrees with the data points at low ${\sqrt{s_{\rm NN}}}$, including our new measurement for $\phi/K^-$.
However calculations from the public $\textup{UrQMD}^{1}$ model~\cite{UrQMD_2:1998,urQMD:1999} underestimate our measurements for both $\phi/K^-$ and $\phi/\Xi^-$. The SMASH (Simulating Many Accelerated Strongly-interacting Hadrons) model~\cite{Elfner_SMASH:2019} attempts to incorporate the newest available experimental data to constrain the resonance branching ratios. These data include both elementary hadronic cross sections and dilepton invariant mass spectra. The $\phi/K^-$ ratio is reasonably reproduced using SMASH in the smaller system and ${\sqrt{s_{\rm NN}}}$ below $\textup{3 GeV}$, despite the overestimation of each individual ($\phi$, $K^-$) transverse mass spectrum measured, e.g. in Au+Au $\textup{0-40\%}$ system by HADES~\cite{HADES_phi_AuAu:2018,Elfner_SMASH:2019}. The predicted $\phi/K^-$ ratio from the same model is about 2.5$\sigma$ higher than central Au+Au 0--10\% collisions at $\textup{3 GeV}$. This indicates that some important in-medium mechanism for strangeness production and propagation may be missing for the large system in SMASH. %Both UrQMD and SMASH transport models highlight the importance of resonance decay contributions in order to reproduce the measured data in the low energies. 
Both UrQMD and SMASH calculations reproduced the measured strangeness data highlighting the importance of the contributions of the resonances in the low energies. 
Furthermore, the $\phi$-meson scattering with the baryonic medium remains an open question from recent measurements of $\pi$ induced nucleus reactions and the $p$-$\phi$ femtoscopy~\cite{HADES_PRL_W_C:2019,ALICE_PRL_pp:2021}. More detailed investigations are needed in order to understand the dynamics of strange and multi-strange hadrons at low energy nuclear collisions.

Our measurement of $K^-$, $\phi$ and $\Xi$ production yields in 3 GeV Au+Au collisions demonstrates the necessity of the Canonical Ensemble for strangeness at low energy heavy-ion collisions. 
%This indicates a different Equation-of-State for the medium created at 3 GeV in comparison to that in higher collision energies. 
In the meantime,  hadronic transport model calculations (UrQMD and SMASH) including resonance contributions reproduce the data. These observations suggest a change of the medium properties at 3 GeV compared to those from higher energy collisions. Similar conclusions have been reached from the measurements of collectivity~\cite{STAR:2021yiu} and high moment of protons~\cite{STAR:2021fge} in 3 GeV Au+Au collisions.

\section{SUMMARY}
\label{summary}

In summary, we report the systematic measurements of $K^-$, $\phi(1020)$ and $\Xi^{-}$ production yields and the $\phi/K^-$, $\phi/\Xi^-$ ratios in Au+Au collisions at ${\sqrt{s_{\rm NN}} = \rm{3\,GeV}}$ with the STAR experiment at RHIC. The measured $\phi/K^-$ ratio is significantly larger than the statistical model prediction based on Grand Canonical Ensemble in the $\textup{0--10\%}$ central collisions. Both the results of $\phi/K^-$ ($r_c \sim 2.7$\,fm) and $\phi/\Xi^-$ ($r_c \sim 4.2$\,fm) ratios favor the Canonical Ensemble model for strangeness production in such collisions. Transport models, including the resonance decays, could reasonably describe our measured $\phi/K^-$ ratio at $\textup{3 GeV}$ and the increasing trend of $\phi/\Xi^-$ at lower energies. 
%Note that the measurement of collective flow from the 3 GeV Au+Au collisions imply a new EoS dominated by baryonic interactions~\cite{STAR:2021yiu}. 
The new results from this paper suggest a significant change in the strangeness production for ${\sqrt{s_{\rm NN}} < \rm{5\,GeV}}$, providing new insights towards the understanding of the QCD medium properties at high baryon density.

% Chapter acknowledgement
\section{Acknowledgement}
\label{acknowledgement}

We would like to thank K. Redlich and J. Steinheimer for fruitful discussions.
We thank the RHIC Operations Group and RCF at BNL, the NERSC Center at LBNL, and the Open Science Grid consortium for providing resources and support.  This work was supported in part by the Office of Nuclear Physics within the U.S. DOE Office of Science, the U.S. National Science Foundation, National Natural Science Foundation of China, Chinese Academy of Science, the Ministry of Science and Technology of China and the Chinese Ministry of Education, the Higher Education Sprout Project by Ministry of Education at NCKU, the National Research Foundation of Korea, Czech Science Foundation and Ministry of Education, Youth and Sports of the Czech Republic, Hungarian National Research, Development and Innovation Office, New National Excellency Programme of the Hungarian Ministry of Human Capacities, Department of Atomic Energy and Department of Science and Technology of the Government of India, the National Science Centre of Poland, the Ministry  of Science, Education and Sports of the Republic of Croatia, German Bundesministerium f\"ur Bildung, Wissenschaft, Forschung and Technologie (BMBF), Helmholtz Association, Ministry of Education, Culture, Sports, Science, and Technology (MEXT) and Japan Society for the Promotion of Science (JSPS).

\bibliography{phi3GeV.bib}

\end{document}